\documentclass[aps,preprint,tightenlines,superscriptaddress]{revtex4}%
\usepackage{amsfonts}
\usepackage{amsmath}
\usepackage{amssymb}
\usepackage{graphicx}%
\providecommand{\U}[1]{\protect\rule{.1in}{.1in}}
\begin{document}
\preprint{ANL-HEP-PR-09-80 and UMTG-11}
\title[EP\&FE]{Evolution profiles and functional equations}
\author{Thomas Curtright$^{\S }$}
\affiliation{Department of Physics, University of Miami, Coral Gables, FL 33124-8046, USA}
\author{Cosmas Zachos$^{\sharp}$}
\affiliation{High Energy Physics Division, Argonne National Laboratory, Argonne, IL
60439-4815, USA\medskip\medskip}
\pacs{PACS number}

\begin{abstract}
Time evolution is formulated and discussed in the framework of Schr\"{o}der's
functional equation. \ The proposed method yields smooth, continuous dynamics
without the prior need for local propagation equations.

\vspace{1in}

\textbf{Contents:}\medskip\medskip

I. \ INTRODUCTION\hfill p 2

II. \ THE METHOD\hfill p 2

III. \ SCHR\"{O}DER'S EXAMPLE\hfill p 4

IV. \ ALL ITERATES OF THE RICKER\ MAP\hfill p 6

V. \ SUMMARY FOR REPULSIVE POLYNOMIAL POTENTIALS\hfill p 15

VI. \ CONCLUSIONS\hfill p 16

Acknowledgments\hfill p 17

References\hfill p 17

\vfill

$^{\S }$curtright@miami.edu$\ \ \ \ \ ^{\sharp}$zachos@anl.gov

\end{abstract}
\volumeyear{year}
\volumenumber{number}
\issuenumber{number}
\eid{identifier}
\startpage{1}
\endpage{ }
\maketitle

\section{Introduction}

From a lattice of time points, as is usually discussed in dynamical systems,
is it possible to obtain continuous, or even smooth, time evolution without
ambiguity? \ Under certain circumstances, the answer is yes, through a
\emph{holographic interpolation process} involving functional methods.

In this paper, we utilize and illustrate Schr\"{o}der's functional techniques
\cite{Schroeder} to produce continuous trajectories out of discrete nonlocal
recursion laws (orbits) for less self-evident dynamical systems, with fixed
points at $x=0$. \ In particular, we build exact continuous iterations of
$2x\left(  1+x\right)  $, and approximations to continuous iterations of
$x\exp x$, and their inverses, and we discuss the notable features of the
resulting evolution.\ \ Typically, smooth, analytic interpolates\ of the
discrete recursion rules that act as the boundary data for the process (hence
our use of the term holographic) are obtained. \ These methods are of possible
use to anyone pursuing practical applications of functional equations
\cite{Castillo}, and are also likely to be relevant to more recent theoretical
developments \cite{Luevano}.

\section{The method}

Consider an evolution trajectory $x\left(  t\right)  $ of a 1-dim system, e.g.
specified by a local, time-translation-invariant law (cf. energy
conservation). \
\begin{equation}
dx\left(  t\right)  /dt=v\left(  x\left(  t\right)  \right)  \ .
\label{LocalTimeEvolution}%
\end{equation}
One may integrate this to obtain the trajectory as a \emph{family of
functions} of the initial data, \emph{indexed by the time},%
\begin{equation}
x\left(  t\right)  =f_{t}\left(  x\left(  0\right)  \right)  \ .
\end{equation}
For any given time, we may scale $t$ to consider a unit time increment $\Delta
t=1$, so that%
\begin{equation}
x\left(  1\right)  =f_{1}\left(  x\left(  0\right)  \right)  \ .
\label{DiscretePropagation}%
\end{equation}
But then, for time-translation-invariant systems, further evolution obeys%
\begin{equation}
x\left(  t+1\right)  =f_{t+1}\left(  x\left(  0\right)  \right)  =f_{1}\left(
x\left(  t\right)  \right)  \ ,
\end{equation}
i.e. $x\left(  t+1\right)  $ is the \emph{same} function of $x\left(
t\right)  $ as $x\left(  1\right)  $ is of $x\left(  0\right)  $. \ For
notational convenience in the following, we will often use $x\left(  0\right)
\equiv x$. \ 

Now suppose, for reasons dictated by physics applications, that the time-local
evolution law (\ref{LocalTimeEvolution}) is not specified, but an explicit,
nonlocal, \emph{discrete propagation function} $f_{1}$ is available, as in
(\ref{DiscretePropagation}). \ We pose the question: \ How does one obtain the
complete, continuous trajectory$\ x\left(  t\right)  =f_{t}\left(  x\right)  $
without benefit of the local relation?

Of course in principle, it is straightforward to compute iterates of
(\ref{DiscretePropagation}) on an integral lattice of time points,
$t=\cdots,-2,-1,0,1,2,3,\cdots$, to obtain \textquotedblleft the splinter of
$x$\textquotedblright\ (for example, see \cite{Small}) , i.e.%
\begin{align}
x(2)  &  =f_{1}\left(  f_{1}\left(  x\right)  \right)  =f_{2}\left(  x\right)
\ ,\nonumber\\
x(n)  &  =f_{1}\left(  f_{1}\cdots\left(  f_{1}\left(  x\right)  \right)
\right)  =f_{n}\left(  x\right)  \ ,\\
x(-1)  &  =f_{1}^{-1}\left(  x\right)  =f_{-1}\left(  x\right)  \ ,\nonumber
\end{align}
etc., assuming the domains for the various functions overlap properly. \ Thus,
$x=f_{-1}\left(  f_{1}\left(  x\right)  \right)  =f_{1}\left(  f_{-1}\left(
x\right)  \right)  $, or more generally, $x(k+n)=f_{k}\left(  f_{n}\left(
x\right)  \right)  =f_{n}\left(  f_{k}\left(  x\right)  \right)  $,
associative and commutative composition. \ From this lattice, upon selecting
derivatives at the lattice points, an infinity of interpolating functions
could be produced, say graphically, to obtain a continuous trajectory. \ This
approach permits easy visualization of the trajectories, but is ambiguous, and
does not place proper emphasis on analytic properties of the solutions, so it
will not be pursued.

Instead, given $f_{1}\left(  x\right)  $ we will use here the theory pioneered
by Ernst Schr\"{o}der \cite{Schroeder} to construct an \emph{analytic}
$f_{t}\left(  x\right)  $ around a fixed point of $f_{1}\left(  x\right)  $.
\ Without loss of generality, we take the fixed point to be $x=0$. \ 

Schr\"{o}der's construction of $f_{t}\left(  x\right)  $ amounts to building
\emph{all} iterates of $f_{1}\left(  x\right)  $, including fractional,
negative, and infinitesimal $t$, based on his eponymous functional conjugacy
equation involving the auxiliary function $\Psi$ \cite{SchroederPun},%
\begin{equation}
s\Psi\left(  x\right)  =\Psi\left(  f_{1}\left(  x\right)  \right)  \ ,
\label{SFE}%
\end{equation}
for some constant $s\neq1$. With the origin a fixed point of $f_{1}$, i.e.
$f_{1}\left(  0\right)  =0$, it follows that $\Psi\left(  0\right)  =0$, and
if $\Psi^{\prime}\left(  0\right)  \neq0$, then $s=f_{1}^{\prime}\left(
0\right)  $. \ The inverse function satisfies Poincar\'{e}'s equation,
\begin{equation}
\Psi^{-1}\left(  sx\right)  =f_{1}\left(  \Psi^{-1}\left(  x\right)  \right)
\ . \label{PFE}%
\end{equation}
Upon iteration of the functional equation, $\Psi$ acts upon the splinter of
$x$\ to give%
\begin{equation}
s^{n}\Psi\left(  x\right)  =\Psi\left(  f_{n}\left(  x\right)  \right)
=\Psi\left(  f_{1}\left(  f_{1}\cdots\left(  f_{1}\left(  x\right)  \right)
\right)  \right)  \ . \label{SFEIterated}%
\end{equation}
Now, the point to be stressed is that \emph{this\ formula naturally yields a
continuous interpolation} for all non-integer $n$,%
\begin{equation}
s^{t}\Psi\left(  x\right)  =\Psi\left(  f_{t}\left(  x\right)  \right)  \ .
\label{SFEInterpolated}%
\end{equation}
So, to produce the full, continuous trajectory, we solve for Schr\"{o}der's
auxiliary function $\Psi\left(  x\right)  $, and construct the inverse
function $\Psi^{-1}$. \ Having done so, this yields $x\left(  t\right)  $ as a
functional similarity transform of the $s^{t}$ multiplicative map. \ From
(\ref{SFEInterpolated}),
\begin{equation}
x\left(  t\right)  \equiv f_{t}\left(  x\right)  =\Psi^{-1}\left(  s^{t}%
~\Psi\left(  x\right)  \right)  \ . \label{FST}%
\end{equation}
In a suitable domain, this gives the \emph{general} iterate for \emph{any} $t
$, analytic around the fixed point $x=0$. \ 

Moreover, this solution manifestly satisfies the requisite associative and
abelian composition properties for all iterates and inverse iterates. \ That
is to say, $f_{t_{1}+t_{2}}\left(  x\right)  =f_{t_{1}}\left(  f_{t_{2}%
}\left(  x\right)  \right)  $, hence $x\left(  t_{1}+t_{2}\right)  =f_{t_{1}%
}\left(  x\left(  t_{2}\right)  \right)  $, as required for
time-translationally invariant systems. \ Some specific cases are as follows.%
\begin{align}
f_{2}\left(  x\right)   &  =\Psi^{-1}\left(  s^{2}~\Psi\left(  x\right)
\right)  =\Psi^{-1}\left(  s\Psi\left(  f_{1}\left(  x\right)  \right)
\right)  =f_{1}\left(  f_{1}\left(  x\right)  \right)  \ ,\nonumber\\
f_{1}\left(  x\right)   &  =\Psi^{-1}\left(  s^{1}~\Psi\left(  x\right)
\right)  =\Psi^{-1}\left(  s^{1/2}~\Psi\left(  \Psi^{-1}\left(  s^{1/2}%
\Psi\left(  x\right)  \right)  \right)  \right)  =f_{1/2}\left(
f_{1/2}\left(  x\right)  \right)  \ ,\\
f_{0}\left(  x\right)   &  \equiv x=\Psi^{-1}\left(  s^{-1}~\Psi\left(
\Psi^{-1}\left(  s^{1}\Psi\left(  x\right)  \right)  \right)  \right)
=f_{-1}\left(  f_{1}\left(  x\right)  \right)  \ ,\nonumber
\end{align}
etc. \ However, it is crucial to note that in the limit $s\rightarrow1$, all
iterates and inverse iterates lose their distinction and degenerate to the
identity map, $f_{0}\left(  x\right)  =x$, and the method fails as written.
\ For this reason, if $f_{1}^{\prime}\left(  0\right)  =1$, one augments
$f_{1}\left(  x\right)  $ in Schr\"{o}der's equation to $sf_{1}\left(
x\right)  $, and takes the $s\rightarrow1$ limit only at the very end of the
calculations, if it makes sense to do so.

\section{Schr\"{o}der's example}

For a very elementary illustration of the technique, consider Schr\"{o}der's
early example of a recursive evolution law,%
\begin{equation}
f_{1}\left(  x\right)  =2x\left(  1+x\right)  \ , \label{f1Example}%
\end{equation}
so $f_{1}\left(  0\right)  =0$, and $s=$ $f_{1}^{\prime}\left(  0\right)  =2$.
\ Schr\"{o}der's equation (\ref{SFE}) is then solved by%
\begin{equation}
\Psi\left(  x\right)  =\frac{1}{2}\ln\left(  1+2x\right)  \ ,\ \ \ \Psi
^{-1}\left(  x\right)  =\frac{1}{2}\left(  e^{2x}-1\right)  \ .
\label{ExamplePsiPsiInverse}%
\end{equation}
This results in%
\begin{equation}
x\left(  t\right)  \equiv f_{t}\left(  x\right)  =\frac{1}{2}\left(  \left(
1+2x\right)  ^{2^{t}}-1\right)  \ , \label{SchroederExample}%
\end{equation}
which indeed obeys $f_{t_{1}}\left(  f_{t_{2}}\left(  x\right)  \right)
=f_{t_{1}+t_{2}}\left(  x\right)  $. \ In particular, $f_{1}\left(  x\right)
=\frac{1}{2}\left(  \left(  1+2x\right)  ^{2}-1\right)  =f_{1/2}\left(
f_{1/2}\left(  x\right)  \right)  $, $f_{1/2}\left(  x\right)  =\frac{1}%
{2}\left(  \left(  1+2x\right)  ^{\sqrt{2}}-1\right)  $, $\ f_{1}^{-1}\left(
x\right)  =f_{-1}\left(  x\right)  =\frac{1}{2}\left(  \sqrt{1+2x}-1\right)
$, etc. \ 

We may visually appreciate the evolution described by (\ref{SchroederExample})
through a surface plot, where the left-right axis is the initial position,
$x$, the depth axis is the time, $t$,\ and the vertical axis is $x\left(
t\right)  $. \ The surface flows from the identity map at $t=0$ to the given
discrete propagation function $x\longmapsto f_{1}\left(  x\right)  $ at $t=1$.
\ The origin is invariant since it is a fixed point of all the functions
$f_{t}\left(  x\right)  $, for every value of the $t$ index, as is the point
$x=-1/2$ in this simple example. \ \ Also note, here we require the initial
$x\geq-1/2$ to have real $x\left(  t\right)  $ for all $t$.%
%TCIMACRO{\FRAME{dtbpFU}{6.2222in}{4.1466in}{0pt}{\Qcb{Figure 1: $\ x\left(
%t\right)  $ plotted versus t and initial $x$ for Schr\"{o}der's example.}}%
%{}{schrodiator__1.eps}{\special{ language "Scientific Word";  type "GRAPHIC";
%width 6.2222in;  height 4.1466in;  depth 0pt;  original-width 6.3135in;
%original-height 4.1981in;  cropleft "0";  croptop "1";  cropright "1";
%cropbottom "0";  filename 'Schrodiator__1.eps';file-properties "XNPEU";}}}%
%BeginExpansion
\begin{center}
\includegraphics[
height=4.1466in,
width=6.2222in
]%
{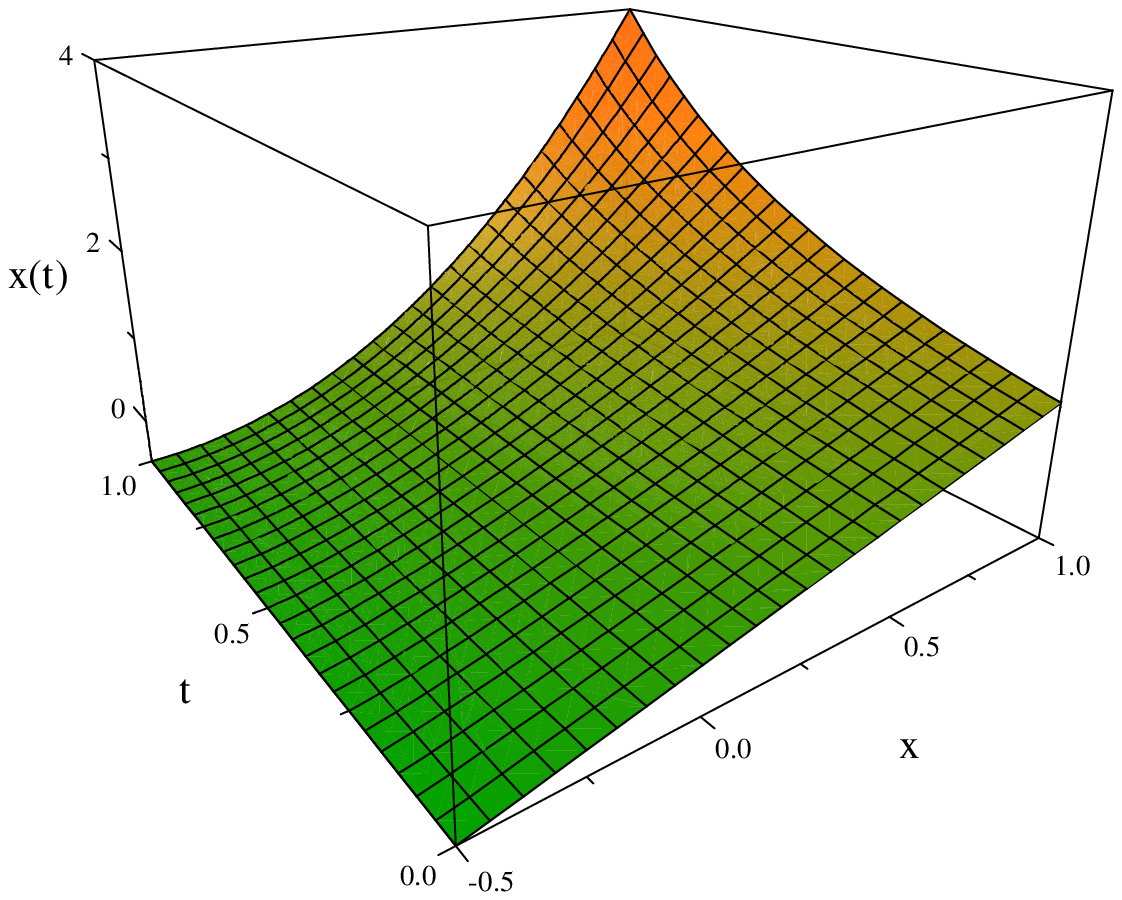}%
\\
Figure 1: $\ x\left(  t\right)  $ plotted versus t and initial $x$ for
Schr\"{o}der's example.
\end{center}
%EndExpansion
The full sweep of the surface describes not just one trajectory, starting at a
single value of $x$ and moving into the page with $t$, but rather it encodes
the evolution of \emph{all} trajectories from the initial domain $x\geq-1/2$.
\ The construction is holographic in the sense that the interior of the
surface is completely determined by the behavior on the time boundaries,
namely, the identity map at the front edge and $f_{1}\left(  x\right)
=2x\left(  1+x\right)  $ at the back edge. \ An animation of sequential, fixed
time, vertical slices through the surface would perhaps show more clearly how
the initial straight line of data evolves to the final $f_{1}$\ \cite{Video}.

The velocity profile following from (\ref{SchroederExample}) is%
\begin{equation}
v\left(  x\left(  t\right)  \right)  =\frac{\partial f_{t}\left(  x\right)
}{\partial t}=\left(  1+2x\left(  t\right)  \right)  \ln\left(  1+2x\left(
t\right)  \right)  \ln\sqrt{2}\ .
\end{equation}%
%TCIMACRO{\FRAME{dtbpFU}{4.1493in}{2.6893in}{0pt}{\Qcb{Figure 2: \ Velocity
%profile for Schr\"{o}der's example.}}{}{schrodiator__2.eps}%
%width 4.1493in;  height 2.6893in;  depth 0pt;  original-width 4.2007in;
%original-height 2.7115in;  cropleft "0";  croptop "1";  cropright "1";
%cropbottom "0";  filename 'Schrodiator__2.eps';file-properties "XNPEU";}}}%
%BeginExpansion
\begin{center}
\includegraphics[
height=2.6893in,
width=4.1493in
]%
{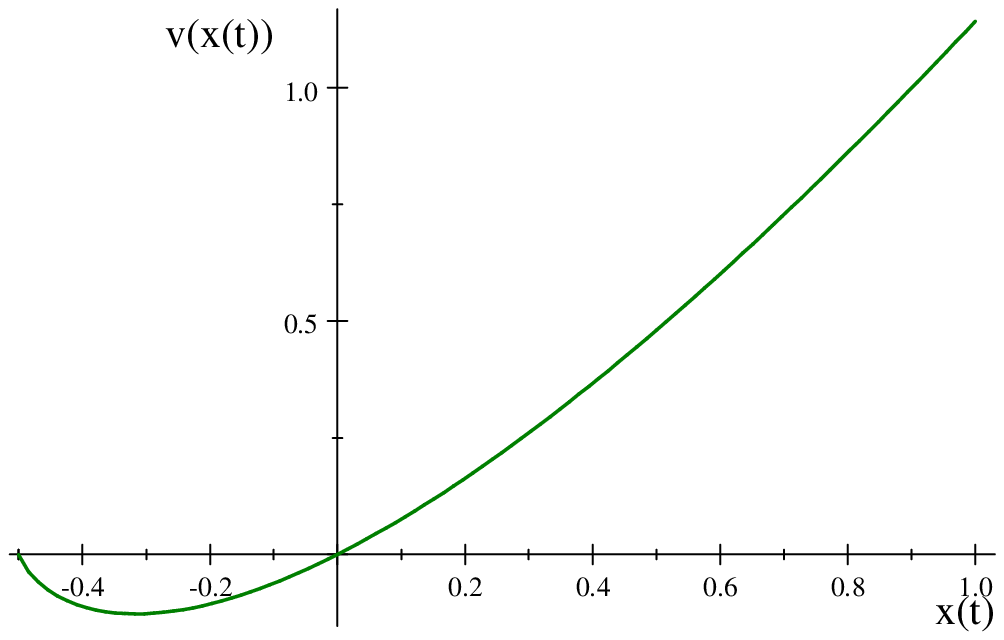}%
\\
Figure 2: \ Velocity profile for Schr\"{o}der's example.
\end{center}
%EndExpansion
Had this velocity been available \emph{ab initio}, it could have been
integrated in the usual way, i.e. $t=\int_{x}^{x\left(  t\right)  }\frac
{ds}{v\left(  s\right)  }$, to obtain the trajectory (\ref{SchroederExample})
moving into the page for each $x$ as shown in Figure 1. \ But here the
starting point was the discrete time step (\ref{f1Example}), with $v\left(
x\left(  t\right)  \right)  $ an emergent feature flowing from the formalism
\cite{SchroederExample}.

Moreover, assuming the motion is governed by a Lagrangian, with $L=\frac{1}%
{2}mv^{2}-V\left(  x\right)  $, and that we are dealing with this system at a
fixed energy, the result for the velocity profile immediately gives the
$x$-dependence of the corresponding potential. \ Namely,
\begin{equation}
V\left(  x\right)  =-\frac{1}{2}mv^{2}\left(  x\right)  +\text{constant .}%
\end{equation}
In this sense the functional method of determining the evolution surface is a
technique to solve \emph{an inverse problem} \cite{Keller,Miller}: \ Given the
$f_{1}\left(  x\right)  $ \textquotedblleft scattering data\textquotedblright%
\ for a finite time step (as opposed to the usual infinite time step in an
idealized scattering process), we may determine an underlying $V\left(
x\right)  $ upon making certain analyticity assumptions about the solution.
\ For the Schr\"{o}der example, the essential features are contained in a plot
of the effective potential $-v^{2}\left(  x\right)  =-\left(  \left(
1+2x\right)  \ln\left(  1+2x\right)  \ln\sqrt{2}\right)  ^{2}$. \
%TCIMACRO{\FRAME{dtbpFU}{4.1493in}{2.6893in}{0pt}{\Qcb{Figure 3: \ Potential
%for Schr\"{o}der's example.}}{}{schrodiator__3.eps}%
%width 4.1493in;  height 2.6893in;  depth 0pt;  original-width 4.2007in;
%original-height 2.7115in;  cropleft "0";  croptop "1";  cropright "1";
%cropbottom "0";  filename 'Schrodiator__3.eps';file-properties "XNPEU";}}}%
%BeginExpansion
\begin{center}
\includegraphics[
height=2.6893in,
width=4.1493in
]%
{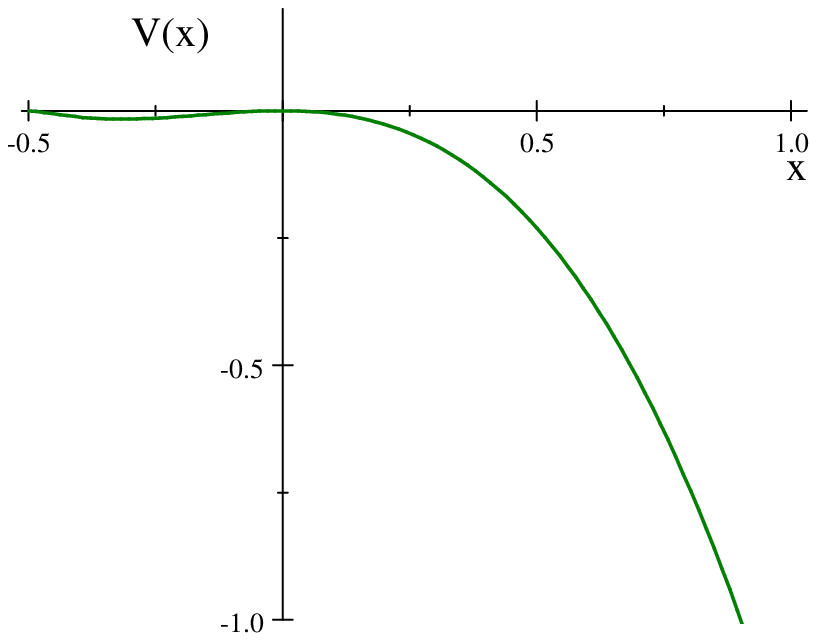}%
\\
Figure 3: \ Potential for Schr\"{o}der's example.
\end{center}
%EndExpansion
Note this $V$ is unbounded below, and more importantly, the aforementioned
fixed points at $x=-1/2$ and $x=0$ are simply \emph{unstable stationary points
of the potential}. \ We submit that this is a common feature of the functional
method, when the results are expressed as a potential function.

In general, a formal solution of Schr\"{o}der's equation is predicated on the
existence of infinite iteration limits of $f_{1}\left(  x\right)  $, as
discovered by Koenigs \cite{Koenigs}. \ (The technique itself is perhaps more
familiar in the context of the Poincar\'{e} equation when written as a
nonlinear finite difference equation.) \ When the splinter is
\textquotedblleft approximately geometric\textquotedblright\ (for example,
again see \cite{Small}) we have%
\begin{equation}
\Psi\left(  x\right)  \propto\lim_{N\rightarrow\infty}\frac{f_{N}\left(
x\right)  }{s^{N}}\ .
\end{equation}
In the above example due to Schr\"{o}der, (\ref{SchroederExample}), we
actually require the limit as $t$ goes to negative infinity,
\begin{equation}
\Psi\left(  x\right)  =\frac{1}{2}\ln\left(  1+2x\right)  =\lim_{N\rightarrow
\infty}\frac{f_{-N}\left(  x\right)  }{2^{-N}}\ ,
\end{equation}
as it is the splinter of $f_{-1}$ that is approximately geometric. \ 

\section{All iterates of the Ricker map}

We now consider the iterates of the Ricker map \cite{Ricker},%
\begin{equation}
f_{1}\left(  x\right)  =x\exp\left(  x\right)  \ , \label{xExp}%
\end{equation}
or equivalently, of its inverse, the Lambert function \cite{Corless},%
\begin{equation}
f_{-1}\left(  x\right)  =\operatorname{LambertW}\left(  x\right)  =\sum
_{n=1}^{\infty}\frac{\left(  -n\right)  ^{n-1}}{n!}x^{n}\ . \label{LambertW}%
\end{equation}
The latter series is convergent for $\left\vert x\right\vert <1/e$. \ Both
$f_{1}\left(  x\right)  $ and $f_{-1}\left(  x\right)  $\ have a unique real
fixed point for $x=0$, so Schr\"{o}der's conjugacy equation can be solved for
the auxiliary $\Psi$ with the same fixed point, in power series expansion
around that point. \ Since $f_{1}^{\prime}\left(  0\right)  =1$, as explained
above, we will modify (\ref{xExp}) and (\ref{LambertW}) to
\begin{equation}
f_{1}\left(  x\right)  =sx\exp\left(  x\right)  \ ,\ \ \ f_{-1}\left(
x\right)  =\operatorname{LambertW}\left(  x/s\right)  \ , \label{xExpMod}%
\end{equation}
thereby introducing a second fixed point in $f_{1}$ at $x=-\ln s$, but
allowing for a possible limit $s\rightarrow1$ at the end of various
calculations, if sensible. \ 

The resulting Schr\"{o}der's equation looks deceptively simple,
\begin{equation}
s\Psi\left(  x\right)  =\Psi\left(  sxe^{x}\right)  \ . \label{xExpSchroeder}%
\end{equation}
Here it is implicit that $\Psi$ really depends on two variables, both $s$ and
$x$, so (\ref{xExpSchroeder}) is actually $s\Psi\left(  x,s\right)
=\Psi\left(  sxe^{x},s\right)  $. \ In the following, this additional
dependence on $s$\ will be understood but usually not displayed. \ It follows
from (\ref{xExpSchroeder}) that $s\Psi^{\prime}\left(  x\right)
=se^{x}\left(  1+x\right)  \Psi^{\prime}\left(  sxe^{x}\right)  $, along with
all higher derivatives with respect to $x$\ given by Fa\`{a} di Bruno's
general formula, hence an explicit series solution for $\Psi$\ about $x=0$ is
straightforward to construct, in principle.

The corresponding Poincar\'{e} equation for the inverse function, $\Phi
\equiv\Psi^{-1}$, while nonlinear, is also relatively simple in appearance
\cite{SchroederPun2}.%
\begin{equation}
\frac{1}{s}~\Phi\left(  sx\right)  =\Phi\left(  x\right)  \exp\left(
\Phi\left(  x\right)  \right)  \ . \label{Poincare}%
\end{equation}
Again, it is implicit that $\Phi$ depends on both $s$ and $x$. \ 

With the normalization choice $\Psi^{\prime}\left(  0\right)  =1$, the
auxiliary function and its inverse are given explicitly to $O\left(
x^{5}\right)  $ by%
\begin{align}
\Psi\left(  x\right)   &  =x-\frac{1}{\left(  s-1\right)  }x^{2}+\frac{1}%
{2}\frac{3s+1}{\left(  s-1\right)  \left(  s^{2}-1\right)  }x^{3}-\frac{1}%
{6}\frac{16s^{3}+8s^{2}+11s+1}{\left(  s-1\right)  \left(  s^{2}-1\right)
\left(  s^{3}-1\right)  }x^{4}\nonumber\\
&  +\frac{1}{24}\frac{125s^{6}+75s^{5}+145s^{4}+146s^{3}+53s^{2}%
+31s+1}{\left(  s-1\right)  \left(  s^{2}-1\right)  \left(  s^{3}-1\right)
\left(  s^{4}-1\right)  }x^{5}+O\left(  x^{6}\right)  \ ,\label{ExplicitPsi}\\
& \nonumber\\
\Psi^{-1}\left(  x\right)   &  =x+\frac{1}{\left(  s-1\right)  }x^{2}+\frac
{1}{2}\frac{3+s}{\left(  s-1\right)  \left(  s^{2}-1\right)  }x^{3}+\frac
{1}{6}\frac{16+11s+8s^{2}+s^{3}}{\left(  s-1\right)  \left(  s^{2}-1\right)
\left(  s^{3}-1\right)  }x^{4}\nonumber\\
&  +\frac{1}{24}\frac{125+131s+145s^{2}+106s^{3}+53s^{4}+15s^{5}+s^{6}%
}{\left(  s-1\right)  \left(  s^{2}-1\right)  \left(  s^{3}-1\right)  \left(
s^{4}-1\right)  }x^{5}+O\left(  x^{6}\right)  \ . \label{ExplicitPsiInverse}%
\end{align}
For purposes of illustration, we plot $O\left(  x^{10}\right)  $
approximations for the auxiliary function and its inverse versus $x$ for
various values of $s$, to obtain the surfaces below. \ We wish to convey only
qualitative behavior at this point, not detailed structure, and to point out
that for values of $s>1$ the auxiliaries have many features similar to those
for Schr\"{o}der's example, (\ref{ExamplePsiPsiInverse}).

\noindent\hspace{-0.75in}%
%TCIMACRO{\FRAME{itbpFU}{3.9258in}{2.6033in}{0in}{\Qcb{Figure 4: $\ \Psi$
%contour surface to $O\left(  x^{10}\right)  $ for $-0.5\leq x\leq0.5$ and
%$2.0\geq s\geq1.5$.}}{}{schrodiator__4.eps}%
%width 3.9258in;  height 2.6033in;  depth 0in;  original-width 3.9719in;
%original-height 2.6246in;  cropleft "0";  croptop "1";  cropright "1";
%cropbottom "0";  filename 'Schrodiator__4.eps';file-properties "XNPEU";}}}%
%BeginExpansion
{\parbox[b]{3.9258in}{\begin{center}
\includegraphics[
height=2.6033in,
width=3.9258in
]%
{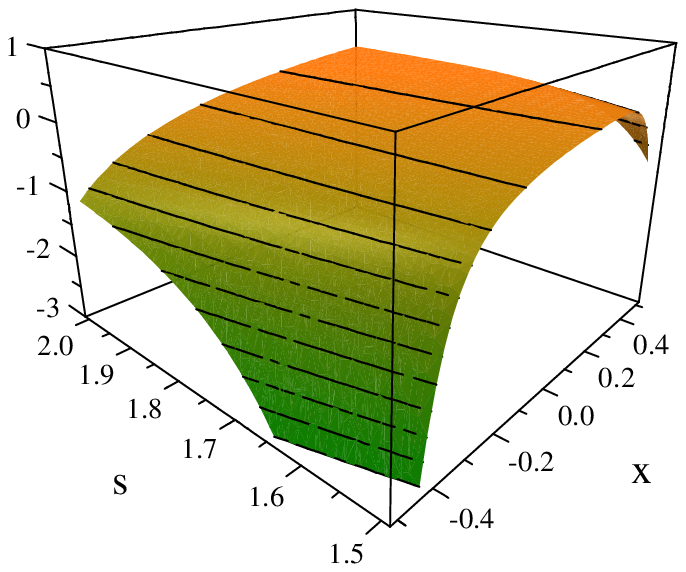}%
\\
Figure 4: $\ \Psi$ contour surface to $O\left(  x^{10}\right)  $ for $-0.5\leq
x\leq0.5$ and $2.0\geq s\geq1.5$.
\end{center}}}%
%EndExpansion
\hspace{-0.2in}%
%TCIMACRO{\FRAME{itbpFU}{3.9258in}{2.6033in}{0in}{\Qcb{Figure 5: $\ \Psi^{-1}$
%contour surface to $O\left(  x^{10}\right)  $ for $-0.5\leq x\leq0.5$ and
%$2.0\geq s\geq1.5$.}}{}{schrodiator__5.eps}%
%width 3.9258in;  height 2.6033in;  depth 0in;  original-width 3.9719in;
%original-height 2.6246in;  cropleft "0";  croptop "1";  cropright "1";
%cropbottom "0";  filename 'Schrodiator__5.eps';file-properties "XNPEU";}}}%
%BeginExpansion
{\parbox[b]{3.9258in}{\begin{center}
\includegraphics[
height=2.6033in,
width=3.9258in
]%
{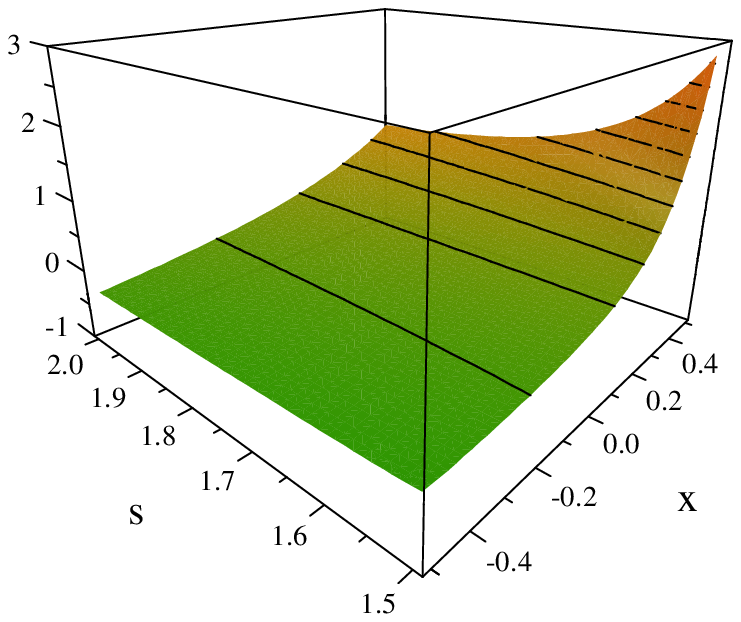}%
\\
Figure 5: $\ \Psi^{-1}$ contour surface to $O\left(  x^{10}\right)  $ for
$-0.5\leq x\leq0.5$ and $2.0\geq s\geq1.5$.
\end{center}}}%
%EndExpansion

\noindent We stress that these are only approximate representations for the
auxiliaries, based on the explicit series to $O\left(  x^{10}\right)  .$ \ We
do not give the $O\left(  x^{10}\right)  $ series explicitly for all $s$,
although they are straightforward to obtain. \ For example, at the back and
front edges of the plotted surfaces, the series are given numerically by%
\begin{align}
\left.  \Psi\left(  x\right)  \right\vert _{s=2}  &  =x-1.0x^{2}%
+1.\,\allowbreak166\,7\allowbreak x^{3}-1.\,\allowbreak452\,4x^{4}%
+1.\,\allowbreak873\,4\allowbreak x^{5}-2.\,\allowbreak470\,8x^{6}%
+3.\,\allowbreak308\,5\allowbreak x^{7}\nonumber\\
&  -4.\,\allowbreak478\,8x^{8}+6.\,\allowbreak113\,3\allowbreak x^{9}%
-8.\,\allowbreak398x^{10}+O\left(  x^{11}\right)  \ ,
\end{align}%
\begin{align}
\left.  \Psi^{-1}\left(  x\right)  \right\vert _{s=2}  &  =\allowbreak
x+x^{2}+0.833\,33x^{3}+\allowbreak0.619\,05x^{4}+0.424\,21\allowbreak
x^{5}+0.273\,6x^{6}+0.168\,26\allowbreak x^{7}\nonumber\\
&  +9.\,\allowbreak952\,9\times10^{-2}x^{8}+5.\,\allowbreak698\,9\times
10^{-2}\allowbreak x^{9}+3.\,\allowbreak173\,4\times10^{-2}x^{10}+O\left(
x^{11}\right)  \ ,
\end{align}%
\begin{align}
\left.  \Psi\left(  x\right)  \right\vert _{s=3/2}  &  =\allowbreak
x-2.0x^{2}+4.\,\allowbreak4\allowbreak x^{3}-10.\,\allowbreak049x^{4}%
+23.\,\allowbreak402\allowbreak x^{5}-55.\,\allowbreak143x^{6}%
+130.\,\allowbreak95\allowbreak x^{7}\nonumber\\
&  -312.\,\allowbreak72x^{8}+749.\,\allowbreak86\allowbreak x^{9}%
-1803.\,\allowbreak8x^{10}+O\left(  x^{11}\right)  \ ,
\end{align}%
\begin{align}
\left.  \Psi^{-1}\left(  x\right)  \right\vert _{s=3/2}  &  =\allowbreak
x+2.0x^{2}+3.\,\allowbreak6\allowbreak x^{3}+6.\,\allowbreak049\,1x^{4}%
+9.\,\allowbreak667\,3\allowbreak x^{5}+14.\,\allowbreak861x^{6}%
+22.\,\allowbreak142\allowbreak x^{7}\nonumber\\
&  +32.\,\allowbreak145x^{8}+45.\,\allowbreak656\allowbreak x^{9}%
+63.\,\allowbreak633x^{10}+O\left(  x^{11}\right)  \ .
\end{align}
Also, we have avoided $s=1$ in the plots of Figures 4 and 5 since the explicit
series results exhibit an expected singular behavior as $s\rightarrow1$ for
each of $\Psi$ and $\Psi^{-1}$, considered separately. \ This behavior is
foreshadowed by the growing coefficients in the numerical series for $\left.
\Psi\left(  x\right)  \right\vert _{s=3/2}$ and $\left.  \Psi^{-1}\left(
x\right)  \right\vert _{s=3/2}$. \ \newpage

However, when $\Psi$ and $\Psi^{-1}$ are \emph{composed} as in (\ref{FST}),
the result for $f_{t}\left(  x\right)  $ is well-behaved even in the limit
$s\rightarrow1$. \ Take $s=e^{\varepsilon}$, and expand in powers of
$\varepsilon$, to find%
\begin{align}
\left.  f_{t}\left(  x\right)  \right\vert _{s=e^{\varepsilon}}  &  =\left.
\Psi^{-1}\left(  s^{t}~\Psi\left(  x\right)  \right)  \right\vert
_{s=e^{\varepsilon}}\label{ExplicitFt}\\
&  =\left(  1+t\varepsilon+O\left(  \varepsilon^{2}\right)  \right)
x\nonumber\\
&  +\left(  t+\frac{1}{2}\left(  -1+3t\right)  t\varepsilon+O\left(
\varepsilon^{2}\right)  \right)  x^{2}\nonumber\\
&  +\left(  \frac{1}{2}\left(  -1+2t\right)  t+\frac{1}{2}\left(
-1+2t\right)  ^{2}t\varepsilon+O\left(  \varepsilon^{2}\right)  \right)
x^{3}\nonumber\\
&  +\left(  \frac{1}{12}\left(  5-15t+12t^{2}\right)  t+\frac{1}{12}\left(
-7+35t-56t^{2}+30t^{3}\right)  t\varepsilon+O\left(  \varepsilon^{2}\right)
\right)  x^{4}\nonumber\\
&  +\left(  \frac{1}{24}\left(  -2+3t\right)  \left(  5-12t+8t^{2}\right)
t+\frac{1}{72}\left(  50-315t+673t^{2}-621t^{3}+216t^{4}\right)
t\varepsilon+O\left(  \varepsilon^{2}\right)  \right)  x^{5}\nonumber\\
&  +O\left(  x^{6}\right)  \ .\nonumber
\end{align}
Despite what one might \emph{naively} expect from the form of $f_{1}$ in
(\ref{xExpMod}), the dependence of $f_{t}$ on $s$ is certainly \emph{not}
multiplicative, a point already borne out by $f_{-1}$ in (\ref{xExpMod}). \ Of
course, as $t\rightarrow0$ all $\varepsilon$ dependence (i.e. all orders in
$\varepsilon$) must disappear to yield the identity map $f_{0}\left(
x\right)  =x$, and indeed (\ref{ExplicitFt}) reduces accordingly.

This last result permits us to compute, at least to $O\left(  x^{5}\right)  $,
the initial velocity profile in the limit $s\rightarrow1$, to find
\begin{equation}
\left.  v\left(  x\right)  \right\vert _{s=1}=\lim_{s\rightarrow
1,\ t\rightarrow0}\frac{\partial f_{t}\left(  x\right)  }{\partial t}%
=x^{2}-\frac{1}{2}x^{3}+\frac{5}{12}x^{4}-\frac{5}{12}x^{5}+O\left(
x^{6}\right)  \ . \label{InitialVelocity1}%
\end{equation}
A better numerical approximation keeps all terms to $O\left(  x^{10}\right)
$, to give%
\begin{align}
\left.  v\left(  x\right)  \right\vert _{s=1}  &  =\allowbreak x^{2}%
-0.5\allowbreak x^{3}+0.416\,67x^{4}-0.416\,67\allowbreak x^{5}+0.445\,83x^{6}%
-0.480\,56\allowbreak x^{7}\nonumber\\
&  +0.501\,12x^{8}-0.491\,63x^{9}+0.452\,15\allowbreak x^{10}+O\left(
x^{11}\right)  \ , \label{BetterInitialVelocity}%
\end{align}
as plotted here.
%TCIMACRO{\FRAME{dtbpFU}{4.1493in}{2.6893in}{0pt}{\Qcb{Figure 6: \ Initial
%velocity profile to $O\left(  x^{10}\right)  $ as $s\rightarrow1$.}}%
%{}{schrodiator__6.eps}{\special{ language "Scientific Word";  type "GRAPHIC";
%width 4.1493in;  height 2.6893in;  depth 0pt;  original-width 4.2007in;
%original-height 2.7115in;  cropleft "0";  croptop "1";  cropright "1";
%cropbottom "0";  filename 'Schrodiator__6.eps';file-properties "XNPEU";}}}%
%BeginExpansion
\begin{center}
\includegraphics[
height=2.6893in,
width=4.1493in
]%
{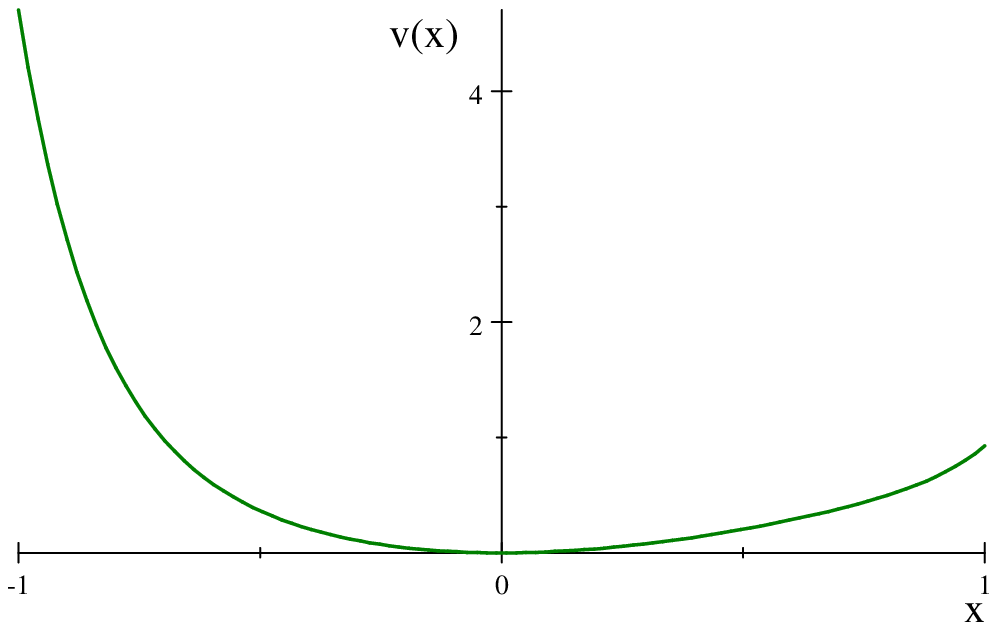}%
\\
Figure 6: \ Initial velocity profile to $O\left(  x^{10}\right)  $ as
$s\rightarrow1$.
\end{center}
%EndExpansion
Moreover, from time-translational invariance, we automatically obtain power
series approximations to $\left.  v\left(  x\left(  t\right)  \right)
\right\vert _{s=1} $ for all $t$ just by substitution of $x\left(  t\right)  $
in either of (\ref{InitialVelocity1}) or (\ref{BetterInitialVelocity}). \ We
may also visualize the effective potential for this problem from a plot of
$-v^{2}\left(  x\right)  $, as discussed previously in the context of
Schr\"{o}der's example.
%TCIMACRO{\FRAME{dtbpFU}{4.1493in}{2.6893in}{0pt}{\Qcb{Figure 7: \ Effective
%potential to $O\left(  x^{10}\right)  $ as $s\rightarrow1$.}}{}%
%{schrodiator__7.eps}{\special{ language "Scientific Word";  type "GRAPHIC";
%width 4.1493in;  height 2.6893in;  depth 0pt;  original-width 4.2007in;
%original-height 2.7115in;  cropleft "0";  croptop "1";  cropright "1";
%cropbottom "0";  filename 'Schrodiator__7.eps';file-properties "XNPEU";}}}%
%BeginExpansion
\begin{center}
\includegraphics[
height=2.6893in,
width=4.1493in
]%
{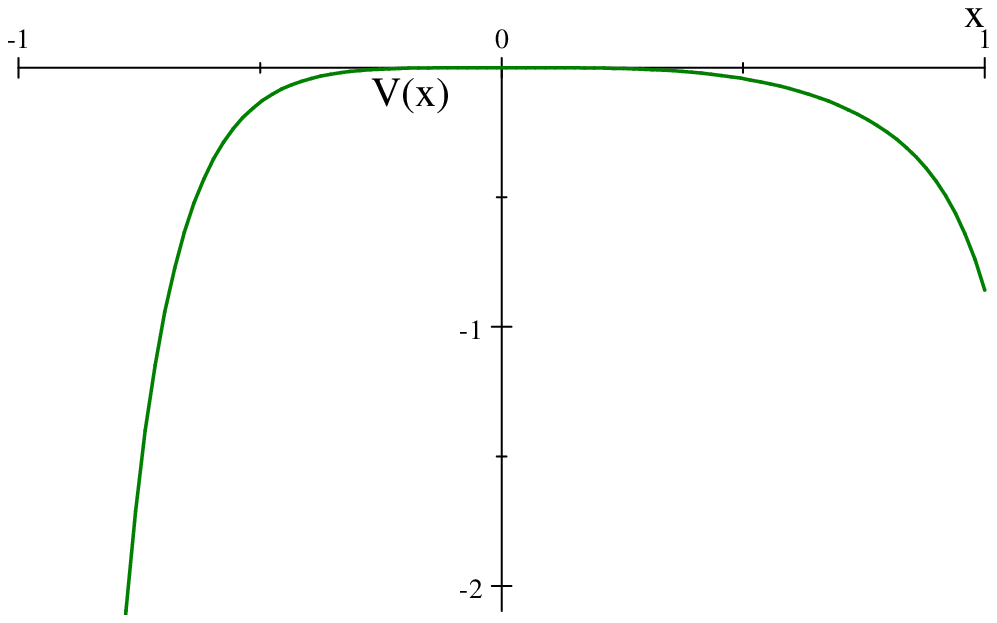}%
\\
Figure 7: \ Effective potential to $O\left(  x^{10}\right)  $ as
$s\rightarrow1$.
\end{center}
%EndExpansion
As in that previous example, we see the fixed point at $x=0$ is a point of
unstable equilibrium.

A similar calculation produces the initial velocity profile for other $s\neq
1$. \ Again to $O\left(  x^{5}\right)  $, we find
\begin{gather}
v\left(  x,s\right)  =\lim_{t\rightarrow0}\frac{df_{t}\left(  x\right)  }%
{dt}=\lim_{t\rightarrow0}\frac{d}{dt}\Psi^{-1}\left(  s^{t}~\Psi\left(
x,s\right)  ,s\right) \label{InitialVelocityLambda}\\
=\left(  \ln s\right)  \left(  x+\frac{1}{s-1}x^{2}-\frac{1}{s^{2}-1}%
x^{3}+\frac{1}{2}\frac{3s+2}{\left(  s^{2}-1\right)  \left(  s^{2}+s+1\right)
}x^{4}-\frac{1}{3}\frac{8s^{2}+4s+3}{\left(  s^{4}-1\right)  \left(
s^{2}+s+1\right)  }x^{5}+O\left(  x^{6}\right)  \right)  \ .\nonumber
\end{gather}
Here we have made explicit the dependence on both $x$ and $s$. \ Of course in
the limit $s\rightarrow1$, $v\left(  x,s\right)  $ remains finite and reduces
to the previous (\ref{InitialVelocity1}). \ We plot the initial velocity
surface versus $x$ and $s$.%
%TCIMACRO{\FRAME{dtbpFU}{4.3586in}{2.8987in}{0pt}{\Qcb{Figure 8: \ Initial
%velocity contour surface, to $O\left(  x^{5}\right)  $, for $-1\leq x\leq1$
%and $1\leq s\leq2$.}}{}{schrodiator__8.eps}%
%width 4.3586in;  height 2.8987in;  depth 0pt;  original-width 4.4136in;
%original-height 2.9253in;  cropleft "0";  croptop "1";  cropright "1";
%cropbottom "0";  filename 'Schrodiator__8.eps';file-properties "XNPEU";}}}%
%BeginExpansion
\begin{center}
\includegraphics[
height=2.8987in,
width=4.3586in
]%
{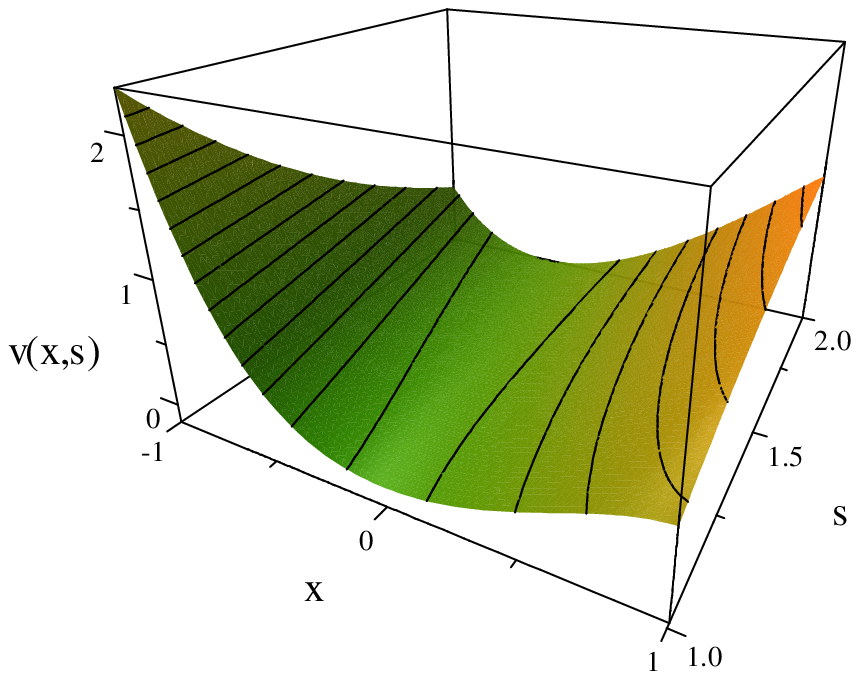}%
\\
Figure 8: \ Initial velocity contour surface, to $O\left(  x^{5}\right)  $,
for $-1\leq x\leq1$ and $1\leq s\leq2$.
\end{center}
%EndExpansion
As before, from time-translational invariance this function automatically also
yields $v\left(  x\left(  t\right)  ,s\right)  $ for all $t$ just by
substitution of $x\left(  t\right)  $ in the power series
(\ref{InitialVelocityLambda}). \ In addition, the potential surface is
effectively given by $V\left(  x,s\right)  =-\frac{1}{2}m\left(  v\left(
x,s\right)  \right)  ^{2}$. \
%TCIMACRO{\FRAME{dtbpFU}{4.3586in}{2.8987in}{0pt}{\Qcb{Figure 9: \ Effective
%potential surface, to $O\left(  x^{5}\right)  $, for $-1\leq x\leq1$ and
%$1\leq s\leq2$.}}{}{schrodiator__9.eps}{\special{ language "Scientific Word";
%type "GRAPHIC";  maintain-aspect-ratio TRUE;  display "USEDEF";
%valid_file "F";  width 4.3586in;  height 2.8987in;  depth 0pt;
%original-width 4.4136in;  original-height 2.9253in;  cropleft "0";
%filename 'Schrodiator__9.eps';file-properties "XNPEU";}}}%
%BeginExpansion
\begin{center}
\includegraphics[
height=2.8987in,
width=4.3586in
]%
{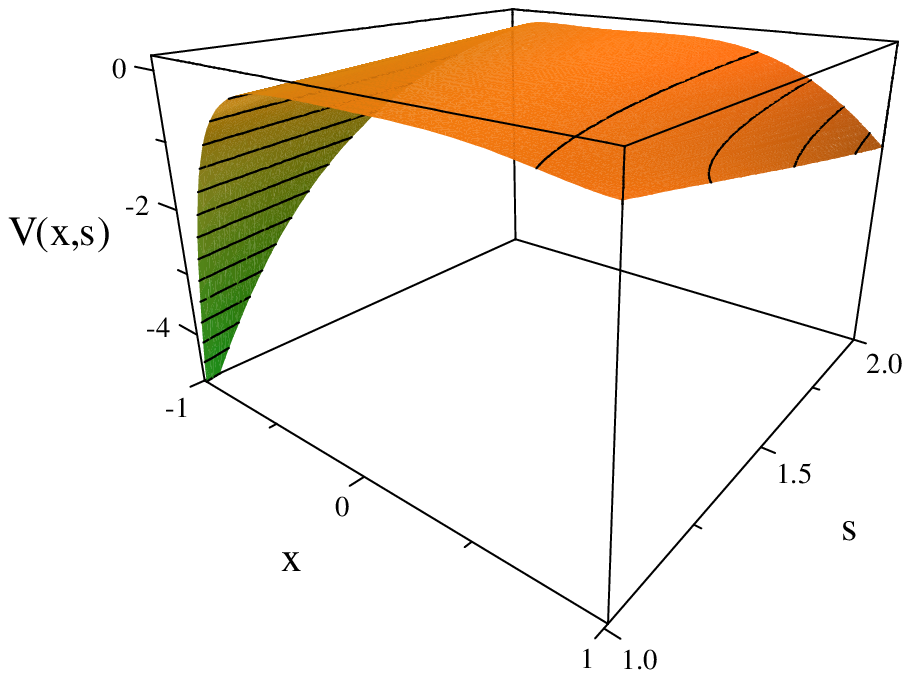}%
\\
Figure 9: \ Effective potential surface, to $O\left(  x^{5}\right)  $, for
$-1\leq x\leq1$ and $1\leq s\leq2$.
\end{center}
%EndExpansion
For fixed $s$ slices, this potential surface again shows unstable equilibria
at the fixed points, $x=0$ and $x=-\ln s$, as was the case for Schr\"{o}der's
elementary example.

To visualize the evolution, we again plot $x\left(  t\right)  $ versus $t$ and
initial $x$, as we did for the closed form results of Schr\"{o}der's simple
example. \ We find the same general features.%
%TCIMACRO{\FRAME{dtbpFU}{6.2222in}{4.1316in}{0pt}{\Qcb{Figure 10: $\ x\left(
%t\right)  =\lim\limits_{s\rightarrow1}f_{t}\left(  x\right)  $, to $O\left(
%x^{10}\right)  $, plotted versus $t$ and initial $x$.}}{}{schrodiator__10.eps}%
%width 6.2222in;  height 4.1316in;  depth 0pt;  original-width 6.3135in;
%original-height 4.183in;  cropleft "0";  croptop "1";  cropright "1";
%cropbottom "0";  filename 'Schrodiator__10.eps';file-properties "XNPEU";}}}%
%BeginExpansion
\begin{center}
\includegraphics[
height=4.1316in,
width=6.2222in
]%
{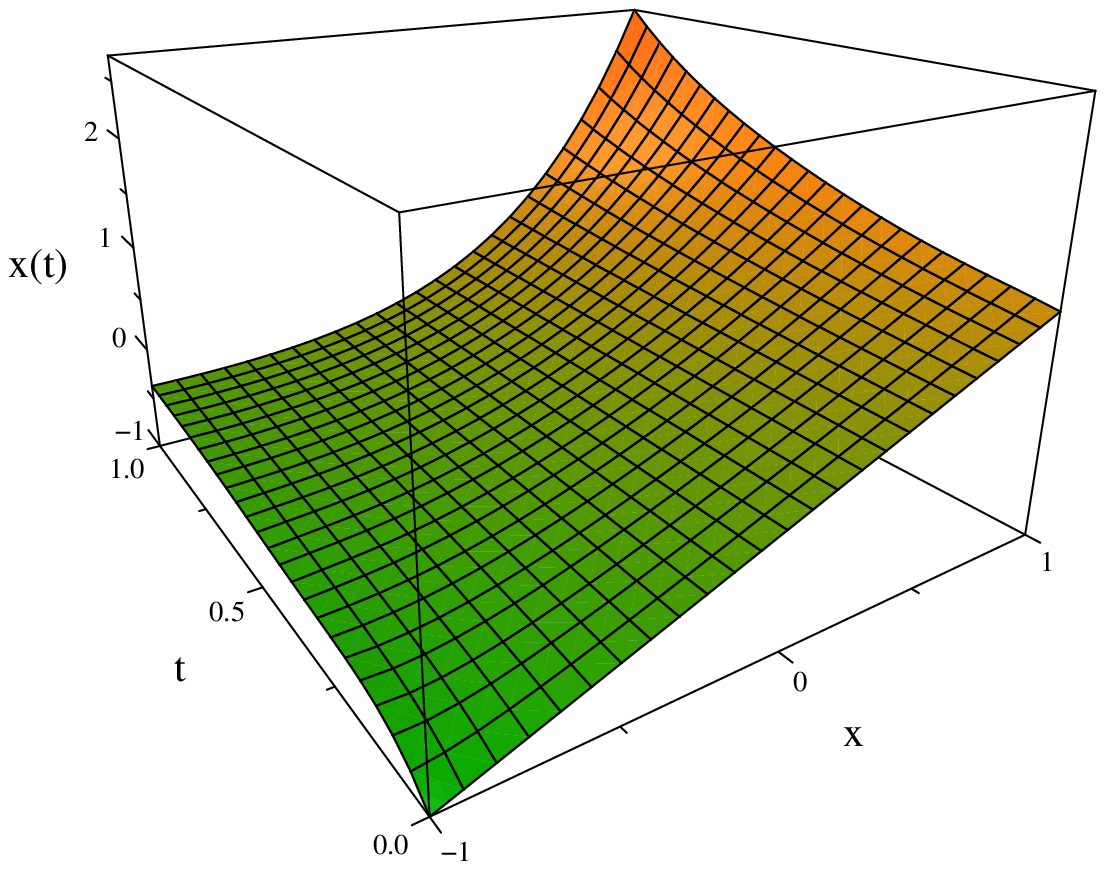}%
\\
Figure 10: $\ x\left(  t\right)  =\lim\limits_{s\rightarrow1}f_{t}\left(
x\right)  $, to $O\left(  x^{10}\right)  $, plotted versus $t$ and initial
$x$.
\end{center}
%EndExpansion
We only display the limit as $s\rightarrow1$. \ Other $s$ give similar surface
geometries. \ As before, the front edge of the surface is the identity map,
and the back edge is the discrete propagation function $f_{1}$. \ Here
$f_{1}\left(  x\right)  =x\exp x$. \ Continuations out of the page would give
the corresponding inverse functions, $f_{-t}$, including $f_{-1}\left(
x\right)  =\operatorname{LambertW}\left(  x\right)  $. \ We reiterate that
this evolution surface was not obtained by the standard method of integrating
velocities for individual initial $x$ to produce the lines that go into the
$t$ depth of the mesh, but rather by the use of functional methods to produce
the continuous, inward flow of complete left to right \textquotedblleft time
slices\textquotedblright\ of the surface \cite{Video}.

With these qualitative images in view, we turn to discuss in more detail the
series solutions for the auxiliaries and the composite functions $f_{t}\left(
x\right)  $ that follow from $f_{1}\left(  x\right)  =sx\exp x$. \ Generally,
the series for $\Psi$ and $\Psi^{-1}$ are of the form%
\begin{equation}
\Psi\left(  x\right)  =\sum_{n=1}^{\infty}\frac{p_{n}\left(  s\right)  ~x^{n}%
}{\left(  n-1\right)  !}\left(
%TCIMACRO{\dprod \limits_{k=1}^{n-1}}%
%BeginExpansion
{\displaystyle\prod\limits_{k=1}^{n-1}}
%EndExpansion
\frac{1}{1-s^{k}}\right)  \ ,\ \ \ \Psi^{-1}\left(  x\right)  =\sum
_{n=1}^{\infty}\frac{q_{n}\left(  s\right)  ~x^{n}}{\left(  n-1\right)
!}\left(
%TCIMACRO{\dprod \limits_{k=1}^{n-1}}%
%BeginExpansion
{\displaystyle\prod\limits_{k=1}^{n-1}}
%EndExpansion
\frac{1}{1-s^{k}}\right)  \ .
\end{equation}
The $p_{n}$ are polynomials in $s$, and can be obtained explicitly by
recursion from $p_{1}=p_{2}=1$ using \cite{ProductConvention}%
\begin{equation}
p_{n}=\left(  n-1\right)  !\sum_{m=1}^{n-1}\frac{p_{m}~s^{m-1}}{\left(
m-1\right)  !}\frac{m^{n-m}}{\left(  n-m\right)  !}\left(
%TCIMACRO{\dprod \limits_{j=m}^{n-2}}%
%BeginExpansion
{\displaystyle\prod\limits_{j=m}^{n-2}}
%EndExpansion
\left(  1-s^{j}\right)  \right)  \text{ \ \ for \ \ }n\geq3\text{ }.
\end{equation}
Direct calculation gives leading and lowest powers of $s$ in each of these
polynomials.%
\begin{gather}
p_{n}=n^{n-2}~s^{\left(  n-1\right)  \left(  n-2\right)  /2}+\left(
n-2\right)  n^{n-3}~s^{n\left(  n-3\right)  /2}+\frac{1}{2}\left(  n-3\right)
\left(  7n-6\right)  n^{n-4}~s^{\frac{1}{2}n\left(  n-3\right)  -1}\nonumber\\
+\frac{1}{6}\left(  n-1\right)  \left(  61n^{2}-338n+384\right)
n^{n-5}~s^{\frac{1}{2}\left(  n+1\right)  \left(  n-4\right)  }\nonumber\\
+\frac{1}{24}\left(  n-1\right)  \left(  705n^{3}-6265n^{2}%
+17\,018n-15\,000\right)  n^{n-6}~s^{\frac{1}{2}\left(  n+1\right)  \left(
n-4\right)  -1}+\cdots\nonumber\\
+\left(  \frac{1}{2}\left(  n-1\right)  \left(  n-2\right)  3^{n-3}-1\right)
s^{2}+\left(  (n-1)\times2^{n-2}-1\right)  s+1\ .
\end{gather}
So written, a power is understood to be \emph{absent }when any exponent in its
coefficient is $<0$. \ There is also an exact \textquotedblleft sum
rule\textquotedblright\ for each polynomial.%
\begin{equation}
p_{n}\left(  s=1\right)  =\left(  ~\left(  n-1\right)  !~\right)  ^{2}\ .
\end{equation}
The auxiliary function's leading asymptotic behavior is given for extreme
values of $s$ by%
\begin{gather}
\Psi\left(  x\right)  ~_{\widetilde{s\rightarrow0}}~\sum_{n=1}^{\infty}%
\frac{x^{n}}{\left(  n-1\right)  !}=xe^{x}\ ,\\
\Psi\left(  x\right)  ~_{\widetilde{s\rightarrow\infty}}~\sum_{n=1}^{\infty
}\left(  -1\right)  ^{n-1}\frac{n^{n-2}s^{\left(  1-n\right)  }x^{n}}{\left(
n-1\right)  !}=s\operatorname{LambertW}\left(  \frac{x}{s}\right)  \ ,
\end{gather}
as well as the formal result,
\begin{equation}
\Psi\left(  x\right)  ~_{\widetilde{s\rightarrow1}}~\sum_{n=1}^{\infty}%
\frac{x^{n}}{\left(  1-s\right)  ^{n-1}}=\frac{x}{1+\frac{x}{s-1}}\ .
\end{equation}
Recall in the power series for $\Psi$ we chose to take the coefficient of
$x$\ to be unity, but that normalization is arbitrary. \ So too are the
normalizations for the asymptotic expressions.

Now consider some particular functional roots and powers. \ An obvious check
on the series for $\Psi$ and $\Psi^{-1}$ is to verify that $\Psi^{-1}\left(
s\Psi\left(  x\right)  \right)  =sxe^{x}$ , and indeed this is true to
$O\left(  x^{5}\right)  $ for the explicit results given in (\ref{ExplicitPsi}%
) and (\ref{ExplicitPsiInverse}). \ Also, when $f_{1}\left(  x\right)
=sxe^{x}$ it is well-known that the inverse function is $f_{-1}\left(
x\right)  =\operatorname{LambertW}\left(  \frac{x}{s}\right)  $, as in
(\ref{xExpMod}). \ We may check that the previous series for $\Psi$ and
$\Psi^{-1}$ do indeed give the series for $\operatorname{LambertW}$ when we
take $t=-1$ in $f_{t}\left(  x\right)  =\Psi^{-1}\left(  s^{t}~\Psi\left(
x\right)  \right)  $. \ For generic $s$, using the explicit series
(\ref{ExplicitPsi}) and (\ref{ExplicitPsiInverse}), we find%
\begin{equation}
\Psi^{-1}\left(  \frac{1}{s}\Psi\left(  x\right)  \right)  =\allowbreak
\frac{1}{s}x+\left(  -\frac{1}{s^{2}}\right)  x^{2}+\frac{3}{2s^{3}}%
x^{3}+\left(  -\frac{8}{3s^{4}}\right)  x^{4}+\frac{125}{24s^{5}}\allowbreak
x^{5}\allowbreak+O\left(  x^{6}\right)  \ ,
\end{equation}
in perfect agreement with the textbook power series for
$\operatorname{LambertW}\left(  \frac{1}{s}x\right)  $. \ As another example,
the functional square-root, such that $f_{1/2}\left(  f_{1/2}\left(  x\right)
\right)  =sxe^{x} $, follows immediately for any $s$. \ For instance,%
\begin{equation}
f_{1/2}\left(  x\right)  =\Psi^{-1}\left(  2\Psi\left(  x\right)  \right)
=\allowbreak2x+\frac{2}{3}x^{2}+\frac{1}{45}x^{3}+\frac{1}{135}x^{4}%
-\frac{389}{137\,700}x^{5}+O\left(  x^{6}\right)  \ ,
\end{equation}
where we have avoided any irrational coefficients by the choice $s=4$. \ This
is quickly checked to satisfy $f_{1/2}\left(  f_{1/2}\left(  x\right)
\right)  =2xe^{x}$, to the order given.

Integer iterates of $\left.  f_{1}\left(  x\right)  \right\vert _{s=1}=xe^{x}$
are easy to construct, without approximation, and plot. \ We have
$f_{1}\left(  x\right)  =xe^{x}$, $f_{2}\left(  x\right)  =xe^{x}e^{xe^{x}}$,
etc. \ In general, we find the form
\begin{equation}
f_{n+1}\left(  x\right)  =a_{1}\exp\left(  a_{1}\left(  1+a_{2}\left(
1+a_{3}\left(  \cdots\left(  1+a_{n-1}\left(  1+a_{n}\right)  \right)
\cdots\right)  \right)  \right)  \right)  \ ,
\end{equation}
where $a_{1}=xe^{x}$, and for $a_{n\geq2}$ there is the recursion relation,
\begin{equation}
a_{k+1}=\exp\left(
%TCIMACRO{\dprod \limits_{j=1}^{k}}%
%BeginExpansion
{\displaystyle\prod\limits_{j=1}^{k}}
%EndExpansion
a_{j}\right)  \ .
\end{equation}
Plotting these reveals an ordered sequence of upward convex functions for
$x\geq-1$, $f_{n+1}\left(  x\right)  >f_{n}\left(  x\right)  $, and confirms
that each has a minimum at $x=-1$, a fact easily established by the chain rule
of differentiation. \ The minima are also ordered, $f_{n+1}\left(  -1\right)
>f_{n}\left(  -1\right)  $, with $f_{1}\left(  -1\right)  =-1/e$, and approach
the $x$ axis as $n$ increases, $\lim_{n\rightarrow\infty}f_{n}\left(
-1\right)  =0$.

We plot four of these integer iterates. \ The curves are just time slices from
a continuation of the $f_{t}\left(  x\right)  $ surface in Figure 10. \ All
other positive iterates are also upward convex and ordered on the domain
$x\geq-1$, $f_{t_{2}}\left(  x\right)  >f_{t_{1}}\left(  x\right)  $ for
$t_{2}>t_{1}$, have $f_{t}^{\prime}\left(  -1\right)  =0$ with minima at
$x=-1$ for $t>0$, and are easily visualized as intercalated between the curves
given in the following Figures. \
%TCIMACRO{\FRAME{dtbpFU}{4.9405in}{7.4107in}{0pt}{\Qcb{Figure 11: \ First four
%integer iterates of $xe^{x}$\ ($f_{1}$,\ $f_{2}$, $f_{3}$, and $f_{4}$)
%plotted alternately in green and orange, above the identity map in black.}}%
%{}{schrodiator__11.eps}{\special{ language "Scientific Word";
%type "GRAPHIC";  maintain-aspect-ratio TRUE;  display "USEDEF";
%valid_file "F";  width 4.9405in;  height 7.4107in;  depth 0pt;
%original-width 5.0327in;  original-height 7.5784in;  cropleft "0";
%filename 'Schrodiator__11.eps';file-properties "XNPEU";}}}%
%BeginExpansion
\begin{center}
\includegraphics[
height=7.4107in,
width=4.9405in
]%
{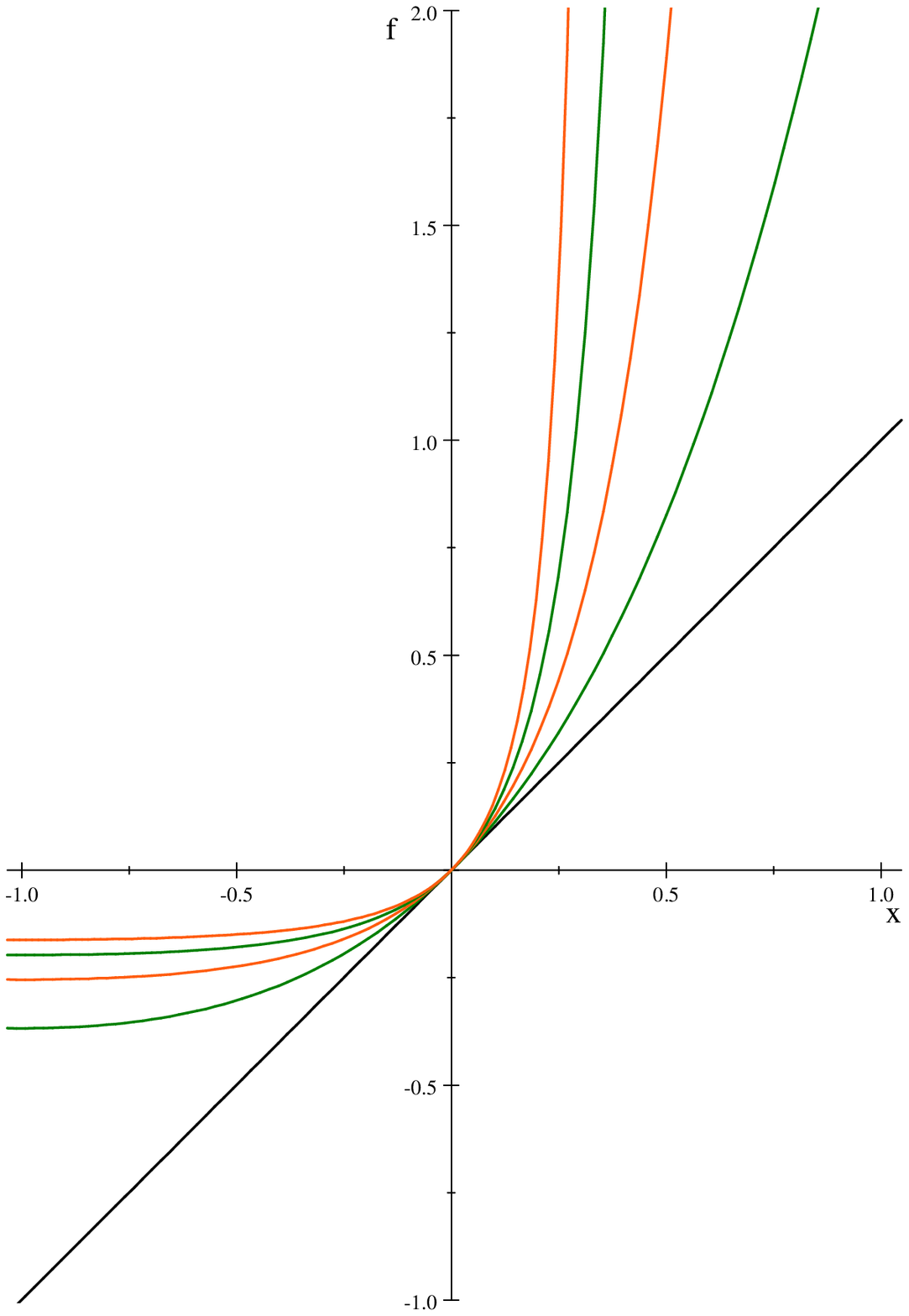}%
\\
Figure 11: \ First four integer iterates of $xe^{x}$\ ($f_{1}$,\ $f_{2}$,
$f_{3}$, and $f_{4}$) plotted alternately in green and orange, above the
identity map in black.
\end{center}
%EndExpansion
Mirror imaging any one of these curves through the straight line of the
identity map in the usual way gives the upper branch of the corresponding
inverse function, e.g. $f_{-1}=\operatorname{LambertW}$, $f_{-2}$, $f_{-3}$,
and $f_{-4}$, and establishes the radius of convergence of the Taylor series
expansion for $f_{-n}\left(  x\right)  $ to be just $R_{n}=\left\vert
f_{n}\left(  -1\right)  \right\vert $. \ 

We also plot some $s\rightarrow1$ series approximations to the fractional
iterates for $-1\leq x\leq0$. \ These again give an ordered sequence of curves
between $f_{1}$ and the identity map. \
%TCIMACRO{\FRAME{dtbpFU}{4.9396in}{4.9396in}{0pt}{\Qcb{Figure 12: \ $O\left(
%x^{10}\right)  $ series approximations for fractional iterates $f_{1}$ (at
%top),\ $f_{1/2}$, $f_{1/4}$, $f_{1/8}$, $f_{1/16}$, and $f_{1/32}$, plotted
%alternately in orange and green, above the identity map in black.}}%
%{}{schrodiator__12.eps}{\special{ language "Scientific Word";
%type "GRAPHIC";  maintain-aspect-ratio TRUE;  display "USEDEF";
%valid_file "F";  width 4.9396in;  height 4.9396in;  depth 0pt;
%original-width 5.0407in;  original-height 5.0407in;  cropleft "0";
%filename 'Schrodiator__12.eps';file-properties "XNPEU";}}}%
%BeginExpansion
\begin{center}
\includegraphics[
height=4.9396in,
width=4.9396in
]%
{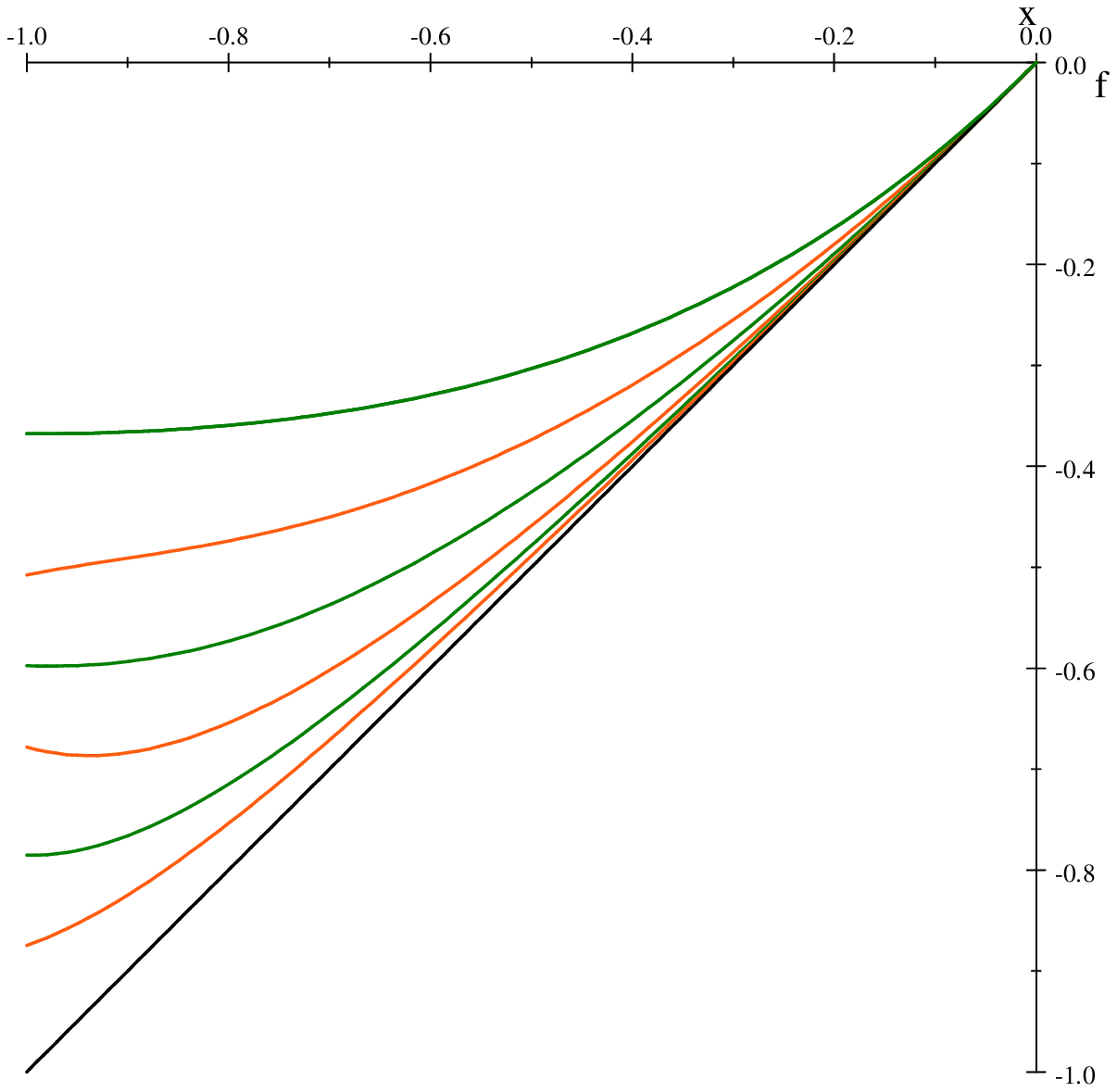}%
\\
Figure 12: \ $O\left(  x^{10}\right)  $ series approximations for fractional
iterates $f_{1}$ (at top),\ $f_{1/2}$, $f_{1/4}$, $f_{1/8}$, $f_{1/16}$, and
$f_{1/32}$, plotted alternately in orange and green, above the identity map in
black.
\end{center}
%EndExpansion
Once more these curves are just time slices from the $f_{t}\left(  x\right)  $
surface in Figure 11. \ For a measure of the accuracy of the $O\left(
x^{10}\right)  $ series approximations used to make this last plot, note that
$f_{t}^{\prime}\left(  -1\right)  $ does not always vanish in the plot, but
for the exact function, \emph{it should}. \ In the plot, disparities are
clearly discernible as $x\rightarrow-1$, especially for the orange curves,
where instead of vanishing we see that $f_{1/2}^{\prime}\left(  -1\right)
>0$, $f_{1/8}^{\prime}\left(  -1\right)  <0$, and $f_{1/32}^{\prime}\left(
-1\right)  >0$. \ Numerically from the series, $\left.  \frac{\partial
}{\partial x}f\left(  x,1\right)  \right\vert _{x=-1}=\allowbreak
-2.\,\allowbreak755\,7\times10^{-6}$,$\left.  \frac{\partial}{\partial
x}f\left(  x,1/2\right)  \right\vert _{x=-1}=\allowbreak0.195\,12$, $\left.
\frac{\partial}{\partial x}f\left(  x,1/4\right)  \right\vert _{x=-1}%
=\allowbreak-0.030\,58$, $\left.  \frac{\partial}{\partial x}f\left(
x,1/8\right)  \right\vert _{x=-1}=\allowbreak-0.291\,32$, $\left.
\frac{\partial}{\partial x}f\left(  x,1/16\right)  \right\vert _{x=-1}%
=\allowbreak-0.030\,586\,$, and $\left.  \frac{\partial}{\partial x}f\left(
x,1/32\right)  \right\vert _{x=-1}=\allowbreak0.346\,87$. \ These numerical
disparities exist because $x=-1$ lies precisely at the radius of convergence
for the power series in question and is a branch point in the exact functions
for generic $t$, so the series become poor approximations as $x\rightarrow-1$.
\ Better numerical results for fractional iterates near $x=-1$ can be obtained
by incorporating branch points into approximate trial solutions of the
functional equation, and then matching these solutions onto the series for
$x>-1$. \ This is work in progress.

\section{Summary for repulsive polynomial potentials}

The method of the paper can be used directly to solve for trajectories in
well-known polynomial potentials, for motion towards or away from unstable
fixed points, when turning points are not encountered in finite time. \ Here
we list the essential features for three examples. \ \bigskip

\textbf{Quadratic:}%
\begin{align}
V\left(  x\right)   &  =-v^{2}\left(  x\right)  =-x^{2}\\
v\left(  x\right)   &  =x\nonumber\\
2\Psi\left(  x\right)   &  =\Psi\left(  2x\right) \nonumber\\
\Psi\left(  x\right)   &  =x\ ,\ \ \ \Psi^{-1}\left(  x\right)  =x\nonumber\\
x\left(  t\right)   &  =xe^{t}\text{\ ,\ \ \ }t=\tau\ln\sqrt{4}\text{ ,
\ \ }x_{0}=0\nonumber
\end{align}

\textbf{Quartic:} \cite{Video}%
\begin{align}
V\left(  x\right)   &  =-v^{2}\left(  x\right)  =-1+2x^{2}-x^{4}\\
v\left(  x\right)   &  =\left(  1-x\right)  \left(  1+x\right) \nonumber\\
\frac{1}{3}\Psi\left(  x\right)   &  =\Psi\left(  \frac{2x+1}{x+2}\right)
\nonumber\\
\Psi\left(  x\right)   &  =2\left(  \frac{x-1}{x+1}\right)  \ ,\ \ \ \Psi
^{-1}\left(  x\right)  =\frac{2+x}{2-x}\nonumber\\
x\left(  t\right)   &  =\frac{x-1+\left(  x+1\right)  e^{2t}}{1-x+\left(
x+1\right)  e^{2t}}\text{ , \ \ }t=\tau\ln\sqrt{3}\text{ , \ \ }%
x_{0}=+1\nonumber
\end{align}

\textbf{Sextic:}%
\begin{align}
V\left(  x\right)   &  =-v^{2}\left(  x\right)  =-x^{2}+2x^{4}-x^{6}\\
v\left(  x\right)   &  =x\left(  1-x\right)  \left(  1+x\right) \nonumber\\
\sqrt{2}\Psi\left(  x\right)   &  =\Psi\left(  \frac{\sqrt{2}x}{\sqrt{1+x^{2}%
}}\right) \nonumber\\
\Psi\left(  x\right)   &  =\frac{x}{\sqrt{1-x^{2}}}\ ,\ \ \ \Psi^{-1}\left(
x\right)  =\frac{x}{\sqrt{1+x^{2}}}\nonumber\\
x\left(  t\right)   &  =\frac{xe^{t}}{\sqrt{1-x^{2}+x^{2}e^{2t}}}\text{ ,
\ \ }t=\tau\ln\sqrt{2}\text{ , \ \ }x_{0}=0\nonumber
\end{align}
To produce these examples, we considered right-moving, zero-energy
configurations. \ For convenience we rescaled $t$, and then we constructed the
\emph{exact} series solution to Schr\"{o}der's equation in the neighborhood of
selected fixed points, $x_{0}$. \ From the auxiliaries, $\Psi$, we then
recovered the continuous time iterates. $\ $That is to say, $x\left(
t\right)  =f_{\tau}\left(  x-x_{0}\right)  =\Psi^{-1}\left(  s^{\tau}%
\Psi\left(  x-x_{0}\right)  \right)  $, with $\tau\propto t$ and $x_{0}$ as
given above.

\section{Conclusions}

In conclusion, we suggest taking a broader perspective and considering other
points of view in dynamics that invoke Schr\"{o}der's functional equation,
$s\circ\chi=\chi\circ f$, or its inverse, the Poincar\'{e} equation,
$\chi^{-1}\circ s=f\circ\chi^{-1}$. \ As usual, $s$ is just the simple
multiplicative map, or change of scale, $s:x\rightarrow sx$, while $f$ is a
less trivial, but given function. \ For example:

\begin{itemize}
\item[i)] \emph{Single trajectory maps:} \ Invert a trajectory function
$x\left(  t\right)  $ to obtain the time $t\left(  x\right)  $, at least for
some interval in $t$, assuming $x\left(  0\right)  =0$. \ Then consider
Schr\"{o}der's functional equation written as $\chi\left(  x;v\right)
=v\chi\left(  t\left(  x\right)  ;v\right)  $. \ Is there a cogent relation
between the parameter $v$ and the \emph{initial velocity}? \ What is the
physical meaning of the resulting function $\chi$? \ 
\end{itemize}

\noindent The answers are straightforward. \ Write the equation to place
emphasis on the time dependence: $\ \chi\left(  x\left(  t\right)  ;v\right)
=v\chi\left(  t;v\right)  $. \ Then analyticity near $t=0$ requires
$\chi\left(  0;v\right)  =0$, and, if $\chi^{\prime}\left(  0;v\right)  \neq
0$, then $v=\left.  \frac{dx\left(  t\right)  }{dt}\right\vert _{t=0}$. \ We
also have $x\left(  t\right)  =\chi^{-1}\left(  v~\chi\left(  t\right)
\right)  $, where additional $v$ dependence of $\chi$, if any, is now
implicit. \ So, dynamical evolution along a single trajectory is in this sense
a functional similarity transformation acting on the initial velocity:
\ $x=\chi^{-1}\circ v\circ\chi$. \ As a consequence of any non-commutativity
between $\chi$ and the simple multiplicative map, we will have $\chi^{-1}\circ
v\circ\chi\neq v$, and the trajectory will evolve as a \emph{non}linear
function of $t$ -- with linear $t$ dependence being just \emph{free} particle
behavior: $x\left(  t\right)  =vt$. \ We may again think of $\chi$ as an
auxiliary function defined on the initial phase space which encodes all
solutions of the classical equations of motion.

\begin{itemize}
\item[ii)] \emph{Pseudo-scaling:} \ Given a trajectory, again construct the
inverse function $x^{-1}$. \ Then what is the significance of the \emph{new
time variable} $T\left(  t;s\right)  =x^{-1}\left(  sx\left(  t\right)
\right)  $?
\end{itemize}

\noindent This change of time variable simply \emph{rescales} the solution.
\ That is to say, $X\left(  T\right)  \equiv x\left(  T\left(  t\right)
\right)  =sx\left(  t\right)  $. \ Note that $T\left(  t\right)  $ differs
from a linear function of $t$ only if the trajectory function fails to commute
with the multiplicative map. \ 

\begin{itemize}
\item[iii)] \emph{Iterative time evolution:} \ From a lattice of time points,
is it \emph{always} possible to use Schr\"{o}der's functional equation to
obtain continuous time evolution?
\end{itemize}

\noindent In this paper we have discussed how, under certain circumstances,
the answer to this last question is affirmative. \ It remains to see whether
the method can be applied in all situations. \ Nevertheless, we believe that
this particular functional method will be applicable to problems in classical
chaos \cite{Cvitanovic,Gilmore}, to complement existing methods for analyzing
unstable fixed points. \ We also think there is no difficulty, in principle,
to prevent the method from being applied to classical dynamics in higher
dimensions, or even to quantum systems, at least under certain circumstances,
upon extension to more variables \cite{Coweny}. \ We look forward to a time
when uses of functional evolution methods have become commonplace.

\begin{acknowledgments}
We thank David Fairlie and Luca Mezincescu for discussions and suggestions.
\ This work was supported in part by the U.S. Department of Energy, Division
of High Energy Physics, under contract DE-AC02-06CH11357, and in part by NSF
Award 0855386.
\end{acknowledgments}


\begin{thebibliography}{99}
\bibitem {SchroederPun}\textquotedblleft Schr\"{o}der's wave
function\textquotedblright\ if you will (pun intended).
\bibitem {SchroederExample}Of course, this elementary example is only an
illustration of the technique. \ A change of variables to $w(t)=\ln(1+2x(t))$
trivializes the problem to merely $dw(t)/dt=\left(  \ln2\right)  ~w(t)$, so
$w(t)=2^{t}~w(0)$, and $w(1)=2$~$w(0)$, hence leading to a monomial auxiliary,
$\Psi=w/2,$ $\Psi^{-1}=2w$. \ Schr\"{o}der generated several such examples by
conformal mappings of such monomial expressions. \ In fact, the change of
variables from $x$ to $w$ is essentially $\Psi(x)=1/2\ln(1+2x)$ above, up to
normalization. Schr\"{o}der's method is summarized by the following
illustrative commutative diagram.%
\[%
\begin{array}
[c]{ccc}%
x & \overset{f_{1}}{\longrightarrow} & f_{1}\left(  x\right) \\
\Psi\left(  x\right)  \ {\Large \downarrow}\ \ \ \ \ \ \ \  &  &
\ \ \ \ \ \ \ \ \ \ \ \ {\Large \downarrow}\ \Psi\left(  f_{1}\left(
x\right)  \right) \\
w & \overset{s}{\longrightarrow} & sw
\end{array}
\]
That is, seek a coordinate transformation $w=\Psi(x)$ s.t. the time-one step
is just $w\mapsto sw$. \ This is easy to iterate for all $t$. \ The over-all
composite map is then $x\mapsto\Psi(f_{1}(x))=s\Psi(x)$.
\bibitem {SchroederPun2}(\ref{Poincare}) amounts to a nonlinear first-order
difference equation in the variables $\ln x$ and $\ln s$, but another way to
describe it is as a nonlocal, nonlinear partial differential equation. \ That
is to say,
\[
s^{x\frac{\partial}{\partial x}-1}~\Phi\left(  x\right)  =\Phi\left(
x\right)  \exp\left(  \Phi\left(  x\right)  \right)  \text{ }.
\]
\textquotedblleft Schr\"{o}der's wave equation\textquotedblright\ if you will
(again, pun intended). \ Various other PDEs follow from taking additional
derivatives of this equation, including partials with respect to $s$, but
suffice it to say that it does not get any better than (\ref{Poincare}). \ A
similar statement applies to Poincar\'{e}'s equation (\ref{PFE}) for any
$f_{1}$, written as $\Phi\left(  sx\right)  =s^{x\partial/\partial x}%
\Phi\left(  x\right)  =f_{1}\left(  \Phi\left(  x\right)  \right)  $.
\bibitem {ProductConvention}We use the convention that the empty product is
unity, e.g. $%
%TCIMACRO{\tprod \limits_{j=n-1}^{n-2}}%
%BeginExpansion
{\textstyle\prod\limits_{j=n-1}^{n-2}}
%EndExpansion
\left(  1-s^{j}\right)  \equiv1$. \ This must be appreciated to understand the
sum rule\ given later in the text.
\end{thebibliography}
\end{document}